\documentclass[12pt]{article}

\usepackage{dcolumn} 
\usepackage{bm} 
\usepackage{a4wide}
\usepackage{fullpage}
\usepackage{slashbox}
\usepackage{tabularx} 
\usepackage{epsf}
\usepackage{amsmath,amsfonts,amssymb}
\usepackage[dvips]{graphicx}
\usepackage{array}
\newcounter{definition}

\def\theequation{\thesection.\arabic{equation}}
\def\appendix{
  \setcounter{section}{0}
  \setcounter{subsection}{0}
  \par
  \def\thesection{Appendix \Alph{section}}
  \def\theequation{\Alph{section}.\arabic{equation}}
  \def\thefigure{\Alph{section}.\arabic{figure}}
}
\newcommand{\nc}{\newcommand}
\nc{\rnc}{\renewcommand}
\nc{\nn}{\nonumber}
\nc{\db}{\displaybreak[0]\\}
\nc\D\Delta
\nc\sg\sigma
\nc\lam\lambda
\nc\kp\kappa
\nc\e\epsilon
\nc\om\omega
\nc{\bra}{\langle}
\nc{\ket}{\rangle}
\nc{\vac}{|0\ket}
\nc{\sh}{\sinh}
\nc{\ch}{\cosh}
\rnc{\(}{\left(}
\rnc{\)}{\right)}
\rnc{\[}{\left[}
\rnc{\]}{\right]}
\nc{\za}[1]{\zeta_a(#1)}
\nc{\zcor}[2]{\bra S_{#1}^z S_{#2}^z \ket}
\nc{\xcor}[2]{\bra S_{#1}^x S_{#2}^x \ket}
\nc{\corf}[8]{\bra S_{#1}^{#5} S_{#2}^{#6} S_{#3}^{#7} S_{#4}^{#8} \ket}
\nc{\cors}[6]{\bra S_1^{#1} S_2^{#2} S_3^{#3} S_4^{#4} S_5^{#5} S_6^{#6} \ket}
\nc\cd\cdots
\nc\rd{P^{\e'_1,\cd,\e'_n}_{\e_1,\cd,\e_n}}
\nc\xab{x_a-x_b}
\nc\bi[2]{\begin{pmatrix}#1\\#2\end{pmatrix}}
\title{Density matrix elements and entanglement entropy 
for the spin-1/2 XXZ chain at $\D$=1/2}
\author{
  Jun Sato           ${}^1$ \thanks{junji@issp.u-tokyo.ac.jp},   \ \
  Masahiro Shiroishi ${}^1$ \thanks{siroisi@issp.u-tokyo.ac.jp}
\\
\\ 
\it ${}^1$
Institute for Solid State Physics, University of Tokyo,\\\it 
  Kashiwanoha 5-1-5, Kashiwa, Chiba 277-8581, Japan\\\it
}
\begin{document}
\maketitle
%
\begin{abstract}
We have analytically obtained all the density matrix elements up to six lattice sites 
for the spin-1/2 Heisenberg XXZ chain at $\D=1/2$. 
We use the multiple integral formula of the correlation function for the massless XXZ chain 
derived by Jimbo and Miwa. 
As for the spin-spin correlation functions, 
we have newly obtained the fourth- and fifth-neighbour transverse correlation functions. 
We have calculated all the eigenvalues of the density matrix 
and analyze the eigenvalue-distribution. 
Using these results the exact values of the entanglement entropy 
for the reduced density matrix up six lattice sites have been obtained. 
We observe that our exact results agree quite well with the asymptotic formula 
predicted by the conformal field theory. 
\end{abstract}
\newpage
\noindent
\section{Introduction}
The spin-1/2 antiferromagnetic Heisenberg XXZ chain is one of the most
fundamental models for one-dimensional quantum magnetism,
which is given by the Hamiltonian
\begin{align}
\mathcal{H}=\sum_{j=-\infty}^{\infty} 
\( S_{j}^x S_{j+1}^x + S_{j}^y S_{j+1}^y + \D S_{j}^z S_{j+1}^z \),
\label{Ham}
\end{align}
where $S_j^{\alpha} = \sg_j^{\alpha}/2$ with $\sg_j^{\alpha}$ being the 
Pauli matrices acting on the $j$-th site and 
$\D$ is the anisotropy parameter. 
For $\D>1$, it is called the massive XXZ model where the system is gapful. 
Meanwhile for $-1<\D\leq1$ case, the system is gapless and  called the massless XXZ model. 
Especially we call it XXX model for the isotropic case $\D=1$. 

The exact eigenvalues and eigenvectors of this model can be obtained 
by the Bethe Ansatz method \cite{Bethe, TakaBook}. 
Many physical quantities in the thermodynamic limit such as specific heat, 
magnetic susceptibility, elementary excitations, etc..., can be exactly evaluated 
even at finite temperature by the Bethe ansatz method \cite{TakaBook}. 

The exact calculation of the correlation functions, however, is still a difficult problem. 
The exceptional case is ${\D=0}$, where the system reduces to a lattice free-fermion model 
by the Jordan-Wigner transformation. 
In this case, we can calculate arbitrary correlation functions 
by means of Wick's theorem \cite{Lieb61,McCoy68}. 
Recently, however, there have been rapid developments 
in the exact evaluations of correlation functions for ${\D\ne0}$ case also, 
since Kyoto Group (Jimbo, Miki, Miwa, Nakayashiki) 
derived a multiple integral representation for arbitrary correlation functions. 
Using the representation theory of the quantum affine algebra $U_q(\hat{sl_2})$, 
they first derived a multiple integral representation for 
massive XXZ antiferromagnetic chain in 1992 \cite{JMMN, JMBook}, 
which is before long extended to the XXX case \cite{Nakayashiki, KIEU} 
and the massless XXZ case \cite{JM}. 
Later the same integral representations were reproduced by 
Kitanine, Maillet, Terras \cite{KMT} in the framework of Quantum Inverse Scattering Method. 
They have also succeeded in generalizing the integral representations to the 
XXZ model with an external magnetic field \cite{KMT}. 
More recently the multiple integral formulas were extended to dynamical correlation 
functions as well as finite temperature correlation functions 
\cite{KMST05a, GKS04, GKS05, Sakai07}. 
In this way it has been established now the correlation functions for XXZ model are 
represented by multiple integrals in general. 
However, these multiple integrals 
are difficult to evaluate both numerically and analytically. 

For general anisotropy $\D$, it has been shown that 
the multiple inetegrals up to four-dimension can be reduced to one-dimensional integrals 
\cite{BK1, BK2, BKNS, SSNT, KSTS03, TKS, KSTS04}. 
As a result all the density matrix elements within {\it four} lattice sites have been 
obtained for general anisotropy \cite{KSTS04}. 
To reduce the multiple integrals into one-dimension, however, involves hard calculation, 
which makes difficult to obtain correlation functions on more than four lattice sites. 
On the other hand, 
at the isotropic point $\D=1$, an algebraic method based on qKZ equation has been 
devised \cite{BKS} and all the density matrix elements up to {\it six} lattice sites 
have been obtained \cite{BST, SST06}. 
Moreover, as for the spin-spin correlation functions, 
up to seventh-neighbour correlation $\zcor{1}{8}$ for XXX chain have been obtained from the 
generating functional approach \cite{SS, SST05}. 
It is desirable that this algebraic method will be generalized to the case with $\D\neq1$. 
Actually, Boos, Jimbo, Miwa, Smirnov and Takeyama have derived an exponential formula 
for the density matrix elements of XXZ model, 
which does not contain multiple integrals \cite{BJMST1, BJMST2, BJMST3, BJMST4, BJMST5}. 
It, however, seems still hard to evaluate the formula for general density matrix elements. 

Among the general $\D\neq0$, there is a special point $\D=1/2$, 
where some intriguing properties have been observed. 
Let us define a correlation function called Emptiness 
Formation Probability (EFP) \cite{KIEU} which signifies the probability to find 
a ferromagnetic string of length $n$: 
\begin{align}
P(n) \equiv \left\bra \prod_{j=1}^{n} \( \frac{1}{2} +S_j^z \) \right\ket.
\end{align}
The explicit general formula for $P(n)$ at $\D=1/2$ was conjectured in \cite{RS} 
\begin{align}
P(n)=2^{-n^2}\prod_{k=0}^{n-1}\frac{(3k+1)!}{(n+k)!},
\end{align}
which is proportional to the number of alternating sign matrix of size $n\times n$. 
Later this conjecture was proved by the explicit evaluation of the multiple integral 
representing the EFP \cite{KMST02}. 
Remarkably, one can also obtain the exact asymptotic behavior as $n\to\infty$ 
from this formula, 
which is the unique valuable example except for the free fermion point $\D=0$. 
Note also that as for the longitudinal two-point correlation functions at $\D=1/2$, 
up to eighth-neighbour correlation function $\zcor19$ have been obtained in \cite{KMST05b} 
by use of the multiple integral representation for the generating function. 
Most outstanding is that all the results are represented by single rational numbers. 
These results motivated us to calculate other correlation functions at $\D=1/2$. 
Actually we have obtained all the density matrix elements up to {\it six} lattice sites 
by the direct evaluation of the multiple integrals. 
All the results can be written by single rational numbers as expected. 
A direct evaluation of the multiple integrals is possible due to the particularity 
of the case for $\D=1/2$ as is explained below. 

\section{Analytical evaluation of multiple integral}
\setcounter{equation}{0}
Here we shall describe how we analytically obtain the density matrix elements at $\D=1/2$ 
from the multiple integral formula. 
Any correlation function can be expressed as a sum of 
density matrix elements $\rd$, 
which are defined by the ground state expectation value 
of the product of elementary matrices: 
\begin{align}
\rd\equiv\bra E_1^{\e'_1\e_1} \cdots E_n^{\e'_n\e_n} \ket, 
\label{dm}
\end{align}
where
$E_j^{\e'_j\e_j}$ are $2 \times 2$ elementary matrices acting on the $j$-th site as 
\begin{align}
E^{++}_j&=\begin{pmatrix}1&0\\0&0\\\end{pmatrix}_{\!\![j]}=\frac12+S_j^z, \quad
E^{--}_j=\begin{pmatrix}0&0\\0&1\\\end{pmatrix}_{\!\![j]}=\frac12-S_j^z, \nn\\
E^{+-}_j&=\begin{pmatrix}0&1\\0&0\\\end{pmatrix}_{\!\![j]}=S_j^+=S_j^x+i S_j^y, \quad
E^{-+}_j=\begin{pmatrix}0&0\\1&0\\\end{pmatrix}_{\!\![j]}=S_j^-=S_j^x-i S_j^y. \nn
\end{align}
The multiple integral formula of the density matrix element for the massless XXZ chain 
reads \cite{JM}
\begin{align}
\rd=&(-\nu)^{-n(n-1)/2}
\int_{-\infty}^\infty\frac{dx_1}{2\pi}\cd\int_{-\infty}^\infty\frac{dx_n}{2\pi} 
\prod_{a>b}\frac{\sh(x_a-x_b)}{\sh[(x_a-x_b-if_{ab}\pi)\nu]} \nn\\
&\times\prod_{k=1}^n\frac{\sh^{y_k-1}\[(x_k+i\pi/2)\nu\]
\sh^{n-y_k}\[(x_k-i\pi/2)\nu\]}{\ch^nx_k}, 
\end{align}
where the parameter $\nu$ is related to the anisotropy as $\D=\cos\pi\nu$ and 
$f_{ab}$ and $y_k$ are determined as 
\begin{align}
&f_{ab}=(1+{\rm sign}[(s'-a+1/2)(s'-b+1/2)])/2, \nn\\
&y_1>y_2>\cd>y_{s'}, \quad \e'_{y_i}=+ \nn\\
&y_{s'+1}>\cd>y_n, \quad \e_{n+1-y_i}=-. 
\end{align} 
In the case of $\D=1/2$, namely $\nu=1/3$, 
the significant simplification occurs 
in the multiple integrals due to the trigonometric identity 
\begin{align}
\sh(x_a\!\!-\!x_b)&=4\sh[(x_a\!\!-\!x_b)/3] 
\sh[(x_a\!\!-\!x_b\!+\!i\pi)/3]\sh[(x_a\!\!-\!x_b\!-\!i\pi)/3]. 
\label{tri}
\end{align}
Actually if we note that the parameter $f_{ab}$ takes the value $0$ or $1$, 
the first factor in the multiple integral at $\nu=1/3$ can be decomposed as 
\begin{align}
\frac{\sh(x_a-x_b)}{\sh[(x_a-x_b-i\pi)/3]}
&=4\sh\(\frac{x_a-x_b}3\)\sh\(\frac{x_a-x_b+i\pi}3\) \nn\\
&=-1+\om e^{\frac23(\xab)}+\om^{-1} e^{-\frac23(\xab)}, \\
\frac{\sh(x_a-x_b)}{\sh[(x_a-x_b)/3]}
&=4\sh\(\frac{x_a-x_b+i\pi}3\)\sh\(\frac{x_a-x_b-i\pi}3\) \nn\\
&=1+e^{\frac23(\xab)}+e^{-\frac23(\xab)},
\end{align}
where $\om=e^{i\pi/3}$. 
Expanding the trigonometoric functions in the second factor 
into exponentials 
\begin{align}
&\sh^{y-1}\[(x+i\pi/2)/3\]\sh^{n-y}\[(x-i\pi/2)/3\] \nn\\
&=2^{1-n}\(\om^{1/2}e^{x/3}-\om^{-1/2}e^{-x/3}\)^{y-1}
\(\om^{-1/2}e^{x/3}-\om^{1/2}e^{-x/3}\)^{n-y} \nn\\
&=2^{1-n}\sum_{l=0}^{y-1}\sum_{m=0}^{n-y}(-1)^{l+m}\bi{y-1}l\bi{n-y}m
\om^{y-l+m-(n+1)/2}e^{\frac13(n-2l-2m-1)x},
\end{align}
we can explicitly evaluate the multiple integral by use of the formula 
\begin{align}
\int_{-\infty}^\infty\frac{e^{\alpha x}dx}{\ch^nx}
=2^{n-1}B\(\frac{n+\alpha}2,\frac{n-\alpha}2\), \quad {\rm Re}(n\pm\alpha)>0, 
\end{align}
where $B(p,q)$ is the beta function defined by
\begin{align}
B(p,q)=\int^1_0t^{p-1}(1-t)^{q-1}dt, \quad {\rm Re}(p),{\rm Re}(q)>0. 
\end{align}
In this way we have succeeded in calculating all the density matrix elements up to 
{\it six} lattice sites. 
All the results are represented by single rational numbers, 
which are presented in Appendix A. 
As for the spin-spin correlation functions, 
we have newly obtained the fourth- and fifth-neighbour 
{\it transverse} two-point correlation function 
\begin{align}
\xcor12&=-\frac5{32}=-0.15625, \nn\\
\xcor13&=\frac{41}{512}=0.080078125, \nn\\
\xcor14&=-\frac{4399}{65536}=-0.0671234130859375, \nn\\
\xcor15&=\frac{1751531}{33554432}=0.0521996915340423583984375, \nn\\
\xcor16&=-\frac{3213760345}{68719476736}=-0.046766368104727007448673248291015625. \nn
\end{align}
The asymptotic formula of the transverse two-point correlation function 
for the massless XXZ chain is established in \cite{LZ, LT} 
\begin{align}
&\xcor{1}{1+n} \sim A_x(\eta) \frac{(-1)^n}{n^{\eta}}
- \tilde{A}_x(\eta) \frac{1}{n^{\eta+\frac{1}{\eta}}} +\cd, \quad \eta=1-\nu, \nn\\
&A_x(\eta)
= \frac1{8 (1-\eta)^2} \[ \frac{\Gamma\(\frac\eta{2-2 \eta}\)}
{2 \sqrt\pi \Gamma\( \frac1{2-2 \eta} \)} \]^{\eta} 
\exp \[ - \int_0^\infty \(\frac{\sh (\eta t)}{\sh (t) \ch[(1-\eta)t]} 
- \eta e^{-2t} \) \frac{dt}t \], \nn\\
&\tilde{A}_x(\eta) = \frac1{2 \eta (1-\eta)} 
\[ \frac{\Gamma\(\frac\eta{2-2 \eta}\)}{2 \sqrt\pi \Gamma\(\frac1{2-2 \eta} \)} \]
^{\eta+\frac{1}{\eta}} 
\exp \[ - \int_0^\infty \(\frac{\ch (2\eta t) e^{-2t}-1}
{2 \sh (\eta t) \sh(t) \cosh[ (1-\eta)t]}
\right.\right.\nn\\&\left.\left.
\hspace{8cm}+\frac{1}{\sh(\eta t)}- \frac{\eta^2+1}{\eta} e^{-2t} \) \frac{dt}t\], 
\end{align}
which produces a good numerical value even for small $n$ as is shown in Table \ref{tr}.
\begin{table}
\begin{center}
\caption{Comparison with the asymptotic formula of the transverse correlation function}
\label{tr}
\begin{tabular}
{@{\hspace{\tabcolsep}\extracolsep{\fill}}c|ccccc} 
\hline
&$\xcor12$&$\xcor13$&$\xcor14$&$\xcor15$&$\xcor16$\\
\hline
Exact
&$-0.156250$&$0.0800781$&$-0.0671234$&$0.0521997$&$-0.0467664$\\
\hline
Asymptotics&$-0.159522$&$0.0787307$&$-0.0667821$&$0.0519121$&$-0.0466083$\\
\hline
\end{tabular}
\end{center}
\end{table}
Note that the {\it longitudinal} correlation function was obtained up to eighth-neighbour 
correlaion $\zcor19$ from the multiple integral representation 
for the generating function \cite{KMST05b}. 
Note also that up to third-neighbour both longitudinal and transverse correlation functions 
for general anisotropy $\D$ were obtained in \cite{KSTS04}. 

\section{Reduced density matrix and entanglement entropy}
\begin{figure}
\includegraphics[width=0.49\textwidth]{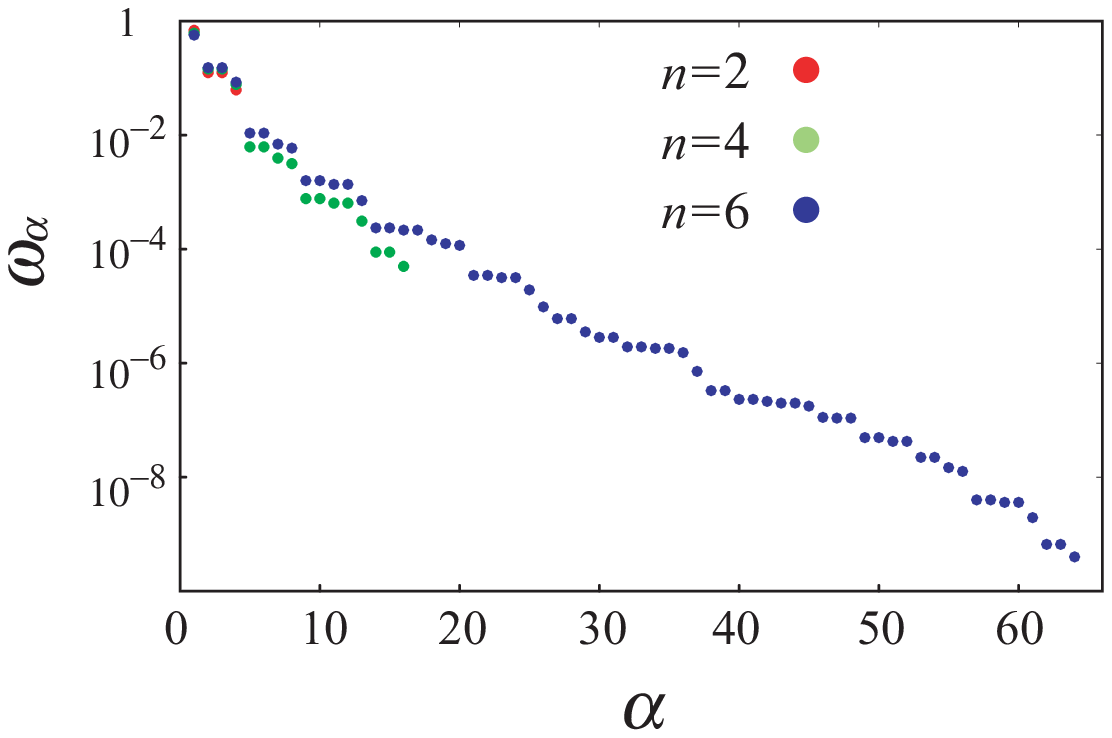}
\includegraphics[width=0.49\textwidth]{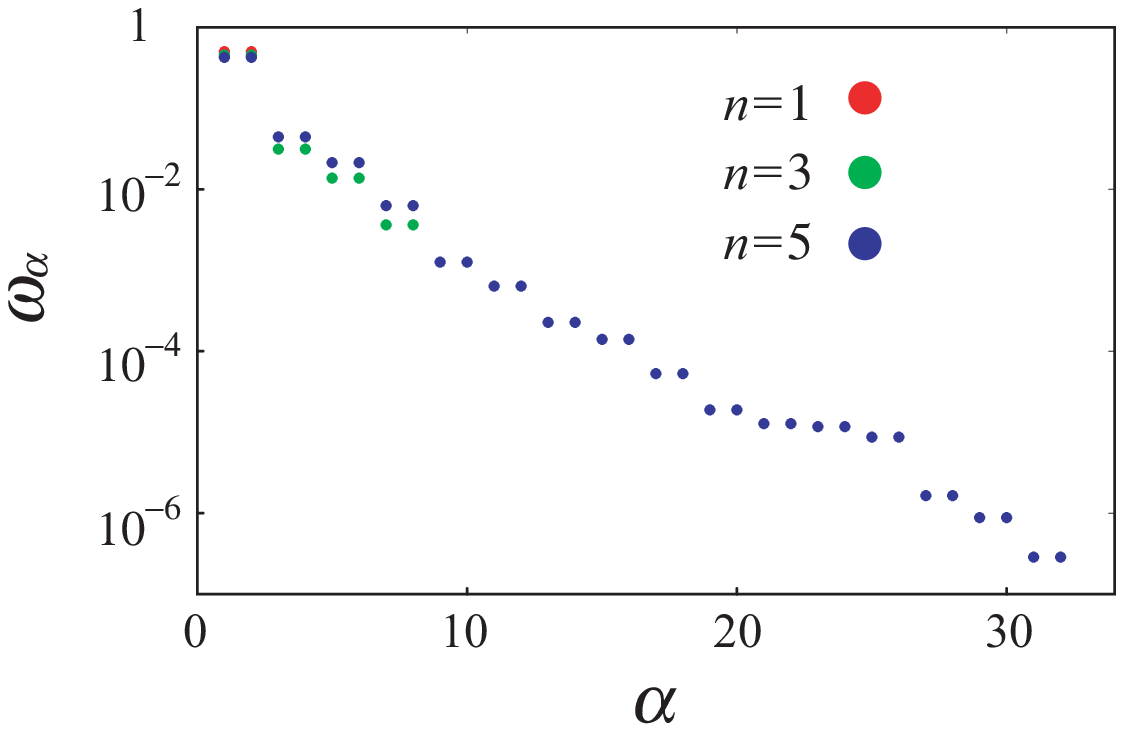}
\caption{Eigenvalue-distribution of density matrices}
\label{ev}
\end{figure}

\setcounter{equation}{0}
Below let us discuss the reduced density matrix for a sub-chain 
and the entanglement entropy. 
The density matrix for the infinite system at zero temperature has the form 
\begin{align}
\rho_{\rm T}\equiv|{\rm GS}\ket\bra{\rm GS}|,
\end{align}
where $|{\rm GS}\ket$ denotes the ground state of the total system. 
We consider a finite sub-chain of length $n$, 
the rest of which is regarded as an environment. 
We define the reduced density matrix for this sub-chain by tracing out the environment 
from the infinite chain
\begin{align}
\rho_n\equiv{\rm tr}_E\rho_{\rm T}=\[\rd\]_{\e_j,\e'_j=\pm}. 
\end{align}
We have numerically evaluate all the eigenvalues 
$\omega_{\alpha}$ $(\alpha=1,2,\cdots,2^n)$ of the reduced density matrix $\rho_n$ 
up to $n=6$. 
We show the distribution of the eigenvalues in Figure \ref{ev}. 
The distribution is less degenerate comapared with the isotropic case $\D=1$ 
shown in \cite{SST06}. 
In the odd $n$ case, all the eigenvalues are two-fold degenerate due to the spin-reverse 
symmetry. 

\begin{figure}
\begin{center}
\includegraphics[width=0.6\textwidth]{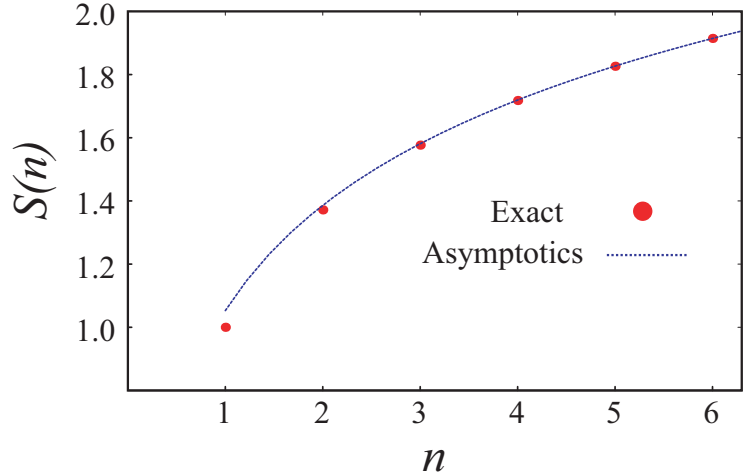}
\caption{Entanglement entropy $S(n)$ of a finite sub-chain of length $n$}
\label{ent}
\end{center}
\end{figure}

Subsequently we exactly evaluate the von Neumann entropy (Entanglement entropy) defined as 
\begin{align}
S(n)\equiv-{\rm tr}\rho_n\log_2\rho_n
=-\sum_{\alpha=1}^{2^n}\omega_{\alpha}\log_2\omega_{\alpha}. 
\end{align}
The exact numerical values of $S(n)$ up to $n=6$ are shown in Table \ref{von}. 
\begin{table}
\begin{center}
\caption{Entanglement entropy $S(n)$ of a finite sub-chain of length $n$}
\label{von}
\begin{tabular}
{@{\hspace{\tabcolsep}\extracolsep{\fill}}cccc} 
\hline
$S$(1)&$S$(2)&$S$(3)&$S$(4)\\
\hline
1&1.3716407621868583&1.5766810784924767&1.7179079372711414\\
\hline
\end{tabular}
\end{center}
\begin{center}
\begin{tabular}
{@{\hspace{\tabcolsep}\extracolsep{\fill}}cc} 
\hline
$S$(5)&$S$(6)\\
\hline
1.8262818282012363&1.9144714710902746\\
\hline
\end{tabular}
\end{center}
\end{table}
By analyzing the behaviour of the entanglement $S(n)$ for large $n$, 
we can see how long quantum correlations reach \cite{VLRK}. 
In the massive region $\D>1$, the entanglement entropy will be saturated as $n$ grows 
due to the finite correlation length. 
This means the ground state is well approximated by a subsystem of a finite length 
corresponding to the large eigenvalues of reduced density matrix. 
On the other hand, in the massless case $-1<\Delta\leq1$, 
the conformal field theory predict that 
the entanglement entropy shows a logarithmic divergence \cite{HLW} 
\begin{align}
S(n)\sim\frac13\log_2n+k_{\D}.
\end{align}
Our exact results up to $n=6$ agree quite well with the asymptotic formula 
as shown in Figure \ref{ent}. 
We estimate the numerical value of the constant term $k_{\D=1/2}$ as 
$k_{\D=1/2}\sim S(6)-\frac13\log_26=1.0528$. 
This numerical value is slightly smaller than the isotropic case $\D=1$, 
where the constant $k_{\D=1}$ is estimated as $k_{\D=1}\sim1.0607$ 
from the exact data for $S(n)$ up to $n=6$ \cite{SST06}. 
At free fermion point $\D=0$, the exact asymptotic formula has been obtained in \cite{JK} 
\begin{align}
&S(n)\sim\frac13\log_2n+k_{\D=0}, \nn\\
&k_{\D=0}=1/3-\int_0^\infty dt
\left\{\frac{e^{-t}}{3t}+\frac1{t\sh^2(t/2)}-\frac{\ch(t/2)}{2\sh^3(t/2)}\right\}/\ln2. 
\end{align}
In this case the numerical value for the constant term is given by 
$k_{\D=0}=1.0474932144\cd$. 

\section{Summary and discussion}
We have succeeded in obtaining 
all the density matrix elements on six lattice sites for XXZ chain at $\D=1/2$. 
Especially we have newly obtained the fourth- and fifth-neighbour {\it transverse} 
spin-spin correlation functions. 
Our exact results for the transverse correlations show good agreement with the 
asymptotic formula established in \cite{LZ, LT}. 
Subsequently we have calculated all the eigenvalues of the reduced density matrix 
$\rho_n$ up to $n=6$. 
From these results we have exactly evaluated the entanglement entropy, 
which shows a good agreement with the asymptotic formula derived via the conformal field 
theory. 
Finally, we remark that similar procedures to evaluate the multiple 
integrals are also possible at $\nu=1/n$ for $n=4,5,6,\cd$, 
since there are similar trigonometric identities as (\ref{tri}). 
We will report the calculation of correlation functions 
for these cases in subsequent papers. 
%
\section*{Acknowledgement}
The authors are grateful to K. Sakai for valuable discussions. 
This work is in part supported by Grant-in-Aid for the Scientific Research (B) No. 18340112. 
from the Ministry of Education, Culture, Sports, Science and Technology, Japan. 
\begin{appendix}
%
\section{Density matrix elements up to $n=6$}
\setcounter{equation}{0}
In this appendix we present all the independent density matrix elements 
defined in eq. (\ref{dm}) up to $n=6$. 
Other elements can be computed from the relations
\begin{align}
&\rd=0 \quad {\rm if} \quad \sum_{j=1}^n\e_j\neq\sum_{j=1}^n\e'_j ,\\
&\rd=P_{\e'_1,\cd,\e'_n}^{\e_1,\cd,\e_n}
=P^{-\e'_1,\cd,-\e'_n}_{-\e_1,\cd,-\e_n}=P^{\e'_n,\cd,\e'_1}_{\e_n,\cd,\e_1}, \\
&P^{+,\e'_1,\cd,\e'_n}_{+,\e_1,\cd,\e_n}+P^{-,\e'_1,\cd,\e'_n}_{-,\e_1,\cd,\e_n}
=P^{\e'_1,\cd,\e'_n,+}_{\e_1,\cd,\e_n,+}+P^{\e'_1,\cd,\e'_n,-}_{\e_1,\cd,\e_n,-}
=P^{\e'_1,\cd,\e'_n}_{\e_1,\cd,\e_n},
\end{align}
and the formula for the EFP \cite{RS, KMST02} 
\begin{align}
P(n)=P^{+,\cd,+}_{+,\cd,+}=2^{-n^2}\prod_{k=0}^{n-1}\frac{(3k+1)!}{(n+k)!}.
\end{align}
\subsection{$n\leq4$}
\begin{align}
P^{-+}_{+-}&=-\frac{5}{16}=-0.3125
, &P^{-++}_{++-}&=\frac{41}{512}=0.0800781
, \nn\db P^{-+++}_{+-++}&=-\frac{221}{8192}=-0.0269775
, &P^{-+++}_{++-+}&=\frac{1579}{65536}=0.0240936
, \nn\db P^{-+++}_{+++-}&=-\frac{289}{32768}=-0.00881958
, &P^{+-++}_{+-++}&=\frac{1037}{16384}=0.0632935
, \nn\db P^{+-++}_{++-+}&=-\frac{2005}{32768}=-0.0611877
, &P^{--++}_{+-+-}&=-\frac{3821}{65536}=-0.0583038
, \nn\db P^{--++}_{++--}&=\frac{1393}{65536}=0.0212555
, &P^{-+-+}_{+-+-}&=\frac{4883}{32768}=0.149017
, \nn\db P^{-++-}_{+--+}&=\frac{3091}{32768}=0.0943298. \nn
\end{align}
\subsection{$n=5$}
\begin{align}
P^{-++++}_{+-+++}&=-\frac{14721}{8388608}=-0.00175488
, & P^{-++++}_{++-++}&=\frac{37335}{16777216}=0.00222534
, \nn\db P^{-++++}_{+++-+}&=-\frac{48987}{33554432}=-0.00145993
, & P^{-++++}_{++++-}&=\frac{13911}{33554432}=0.00041458
, \nn\db P^{+-+++}_{+-+++}&=\frac{179699}{33554432}=0.00535545
, & P^{+-+++}_{++-++}&=-\frac{120337}{16777216}=-0.00717264
, \nn\db P^{+-+++}_{+++-+}&=\frac{165155}{33554432}=0.004922
, & P^{++-++}_{++-++}&=\frac{168313}{16777216}=0.0100322
, \nn\db P^{--+++}_{+--++}&=\frac{31069}{2097152}=0.0148149
, & P^{--+++}_{+-+-+}&=-\frac{411583}{16777216}=-0.0245323
, \nn\db P^{--+++}_{+-++-}&=\frac{196569}{16777216}=0.0117164
, & P^{--+++}_{++-+-}&=-\frac{281271}{33554432}=-0.00838253
, \nn\db P^{--+++}_{+++--}&=\frac{79673}{33554432}=0.00237444
, & P^{-+-++}_{+--++}&=-\frac{1441787}{33554432}=-0.0429686
, \nn\db P^{-+-++}_{+-++-}&=-\frac{1261655}{33554432}=-0.0376002
, & P^{-+-++}_{++-+-}&=\frac{59459}{2097152}=0.0283523
, \nn\db P^{-++-+}_{+-++-}&=\frac{1575515}{33554432}=0.046954
, & P^{-+++-}_{+--++}&=-\frac{696151}{33554432}=-0.0207469
, \nn\db P^{-+++-}_{+-+-+}&=\frac{1366619}{33554432}=0.0407284. \nn
\end{align}
\subsection{$n=6$}
\begin{align}
P^{-+++++}_{+-++++}&=-\frac{1546981}{34359738368}=-0.0000450231
, &P^{-+++++}_{++-+++}&=\frac{5095899}{68719476736}=0.0000741551
, \nn\db P^{-+++++}_{+++-++}&=-\frac{2366275}{34359738368}=-0.0000688677
, &P^{-+++++}_{++++-+}&=\frac{2455833}{68719476736}=0.0000357371
, \nn\db P^{-+++++}_{+++++-}&=-\frac{284577}{34359738368}=-8.28228\times 10^{-6}
, &P^{+-++++}_{+-++++}&=\frac{2927709}{17179869184}=0.000170415
, \nn\db P^{+-++++}_{++-+++}&=-\frac{20086627}{68719476736}=-0.000292299
, &P^{+-++++}_{+++-++}&=\frac{19268565}{68719476736}=0.000280395
, \nn\db P^{+-++++}_{++++-+}&=-\frac{10295153}{68719476736}=-0.000149814
, &P^{++-+++}_{++-+++}&=\frac{17781349}{34359738368}=0.000517505
, \nn\db P^{++-+++}_{+++-++}&=-\frac{35087523}{68719476736}=-0.000510591
, &P^{--++++}_{+--+++}&=\frac{48421023}{34359738368}=0.00140924
, \nn\db P^{--++++}_{+-+-++}&=-\frac{214080091}{68719476736}=-0.00311528
, &P^{--++++}_{+-++-+}&=\frac{88171589}{34359738368}=0.00256613
, \nn\db P^{--++++}_{+-+++-}&=-\frac{57522267}{68719476736}=-0.000837059
, &P^{--++++}_{++--++}&=\frac{56776545}{34359738368}=0.00165241
, \nn\db P^{--++++}_{++-+-+}&=-\frac{154538459}{68719476736}=-0.00224883
, &P^{--++++}_{++-++-}&=\frac{60809571}{68719476736}=0.000884896
, \nn\db P^{--++++}_{+++--+}&=\frac{6708473}{8589934592}=0.000780969
, &P^{--++++}_{+++-+-}&=-\frac{33366621}{68719476736}=-0.000485548
, \nn\db P^{--++++}_{++++--}&=\frac{3860673}{34359738368}=0.00011236
, &P^{-+-+++}_{+--+++}&=-\frac{85706851}{17179869184}=-0.0049888
, \nn\db P^{-+-+++}_{+-+-++}&=\frac{12211375}{1073741824}=0.0113727
, &P^{-+-+++}_{+-++-+}&=-\frac{332557469}{34359738368}=-0.0096787
, \nn\db P^{-+-+++}_{+-+++-}&=\frac{56183761}{17179869184}=0.00327033
, &P^{-+-+++}_{++--++}&=-\frac{430452959}{68719476736}=-0.00626391
, \nn\db P^{-+-+++}_{++-+-+}&=\frac{606065059}{68719476736}=0.00881941
, &P^{-+-+++}_{++-++-}&=-\frac{123612511}{34359738368}=-0.0035976
, \nn\db P^{-+-+++}_{+++--+}&=-\frac{108202041}{34359738368}=-0.00314909
, &P^{-+-+++}_{+++-+-}&=\frac{70061315}{34359738368}=0.00203905
, \nn\db P^{-++-++}_{+--+++}&=\frac{7860495}{1073741824}=0.00732066
, &P^{-++-++}_{+-+-++}&=-\frac{591759525}{34359738368}=-0.0172225
, \nn\db P^{-++-++}_{+-++-+}&=\frac{1044016671}{68719476736}=0.0151924
, &P^{-++-++}_{+-+++-}&=-\frac{367905053}{68719476736}=-0.00535372
, \nn\db P^{-++-++}_{++--++}&=\frac{676957849}{68719476736}=0.00985103
, &P^{-++-++}_{++-+-+}&=-\frac{988973861}{68719476736}=-0.0143915
, \nn\db P^{-++-++}_{++-++-}&=\frac{6581795}{1073741824}=0.00612977
, &P^{-++-++}_{+++--+}&=\frac{363618785}{68719476736}=0.00529135
, \nn\db P^{-+++-+}_{+--+++}&=-\frac{185522333}{34359738368}=-0.00539941
, &P^{-+++-+}_{+-+-++}&=\frac{901633567}{68719476736}=0.0131205
, \nn\db P^{-+++-+}_{+-++-+}&=-\frac{103539423}{8589934592}=-0.0120536
, &P^{-+++-+}_{+-+++-}&=\frac{38524625}{8589934592}=0.00448486
, \nn\db P^{-+++-+}_{++--++}&=-\frac{267901987}{34359738368}=-0.00779697
, &P^{-+++-+}_{++-+-+}&=\frac{12750645}{1073741824}=0.011875
, \nn\db P^{-+++-+}_{+++--+}&=-\frac{309855965}{68719476736}=-0.004509
, &P^{-++++-}_{+--+++}&=\frac{29410257}{17179869184}=0.0017119
, \nn\db P^{-++++-}_{+-+-++}&=-\frac{296882461}{68719476736}=-0.00432021
, &P^{-++++-}_{+-++-+}&=\frac{35985105}{8589934592}=0.00418922
, \nn\db P^{-++++-}_{++--++}&=\frac{92176287}{34359738368}=0.00268268
, &P^{+--+++}_{+--+++}&=\frac{202646807}{34359738368}=0.0058978
, \nn\db P^{+--+++}_{+-+-++}&=-\frac{972245985}{68719476736}=-0.014148
, &P^{+--+++}_{+-++-+}&=\frac{217687057}{17179869184}=0.0126711
, \nn\db P^{+--+++}_{++-+-+}&=-\frac{211696415}{17179869184}=-0.0123224
, &P^{+--+++}_{+++--+}&=\frac{78922695}{17179869184}=0.00459391
, \nn\db P^{+-+-++}_{+-+-++}&=\frac{1196499417}{34359738368}=0.0348227
, &P^{+-+-++}_{+-++-+}&=-\frac{2209522727}{68719476736}=-0.0321528
, \nn\db P^{+-+-++}_{++-+-+}&=\frac{1108384987}{34359738368}=0.0322582
, &P^{+-++-+}_{+-++-+}&=\frac{530683585}{17179869184}=0.0308899
, \nn\db P^{+-++-+}_{++--++}&=\frac{347202525}{17179869184}=0.0202098
, &P^{---+++}_{+--++-}&=-\frac{268623007}{68719476736}=-0.00390898
, \nn\db P^{---+++}_{+-+-+-}&=\frac{46285135}{8589934592}=0.0053883
, &P^{---+++}_{+-++--}&=-\frac{136974885}{68719476736}=-0.00199325
, \nn\db P^{---+++}_{++-+--}&=\frac{19939391}{17179869184}=0.00116063
, &P^{---+++}_{+++---}&=-\frac{18442085}{68719476736}=-0.000268368
, \nn\db P^{--+-++}_{+--++-}&=\frac{1018463205}{68719476736}=0.0148206
, &P^{--+-++}_{+-+-+-}&=-\frac{1454513249}{68719476736}=-0.021166
, \nn\db P^{--+-++}_{+-++--}&=\frac{277721503}{34359738368}=0.00808276
, &P^{--+-++}_{++-+--}&=-\frac{335265249}{68719476736}=-0.00487875
, \nn\db P^{--++-+}_{+--++-}&=-\frac{369408975}{17179869184}=-0.0215024
, &P^{--++-+}_{+-+-+-}&=\frac{1104236607}{34359738368}=0.0321375
, \nn\db P^{--++-+}_{+-++--}&=-\frac{880560357}{68719476736}=-0.0128138
, &P^{--++-+}_{++--+-}&=-\frac{876924641}{68719476736}=-0.0127609
, \nn\db P^{--+++-}_{+---++}&=\frac{113631201}{17179869184}=0.00661421
, &P^{--+++-}_{+--+-+}&=-\frac{292857807}{17179869184}=-0.0170466
, \nn\db P^{--+++-}_{+-+--+}&=\frac{548645951}{34359738368}=0.0159677
, &P^{--+++-}_{++---+}&=-\frac{377925345}{68719476736}=-0.00549954
, \nn\db P^{-+-+-+}_{+--++-}&=\frac{1719255909}{34359738368}=0.0500369
, &P^{-+-+-+}_{+-+-+-}&=-\frac{5350158879}{68719476736}=-0.0778551
, \nn\db P^{-+-++-}_{+--+-+}&=\frac{1565770597}{34359738368}=0.0455699
, &P^{-+-++-}_{+-+--+}&=-\frac{3059753503}{68719476736}=-0.0445253
, \nn\db P^{-++--+}_{+--++-}&=-\frac{2117554719}{68719476736}=-0.0308145. \nn
\end{align}
\end{appendix}

\end{document}